\documentclass[proof]{pasj00}
\begin{document}
\SetRunningHead{M. Ohno, Y. Fukazawa, and N. Iyomoto}{Hard X-Ray Emission from Mrk 1210}
\Received{2003/12/8}%{yyyy/mm/dd}
\Accepted{2004/03/12}%{yyyy/mm/dd}

\title{\bf Reflection Component in the Hard X-Ray Emission from the
 Seyfert 2 Galaxy Mrk 1210}

%%% begin:list of authors
\author{Masanori \textsc{Ohno},  Yasushi \textsc{Fukazawa}}%

\affil{Department of Physical Sciences, School of Science, Hiroshima
 University \\ 1-3-1 Kagamiyama, Higashi-Hiroshima, Hiroshima 739-8526}
\email{ohno@hirax7.hepl.hiroshima-u.ac.jp}
\and
\author{Naoko \textsc{Iyomoto}}
\affil{Laboratory for High Energy Astrophysics, Code 662, \\
  NASA Goddard Space Flight Center, Greenbelt, MD 20771, USA}

 %Do NOT move this preamble from here!

\maketitle

\begin{abstract}

The Seyfert 2 galaxy Mrk 1210 was found to exhibit
 a flat hard X-ray component by ASCA, although ASCA could not distinguish
 whether it is an absorbed direct component or a reflected one.
We then observed Mrk 1210 with BeppoSAX, and found that the X-ray
 spectral properties are quite different from those of ASCA, as have
 been confirmed with XMM-Newton; 
 the flux is significantly higher than that in the ASCA observation, 
 and a clear absorption cut-off appears below 5 keV. 
A bright hard X-ray emission is detected up to 100 keV. 
The reflection component is necessary to describe the BeppoSAX PDS
 spectrum, and represents the ASCA hard component very well.
Therefore, the hard component in the ASCA spectrum is a reflected one, whose intensity is
 almost constant over 6 years.
This indicates that a dramatic spectral variability is attributed 
 to a large change of the absorption column density by a factor of $>5$,
 rather than the variability of the nuclear emission.
The change in the absorption-column density means that  
 the torus is not homogeneous, but has a blobby structure  with a typical
 blob size of $<$ 0.001
 pc. 

\end{abstract}

\section{Introduction}

Seyfert galaxies are classified into type 1 and  2, according to
whether they have broad emission lines in their optical/UV spectra or
not. However, the detection of polarized broad lines in the optical
spectra of the type-2 Seyfert galaxy (\cite{Antou}) demonstrated 
the presence of broad emission-line regions even in Seyfert 2 galaxies. 
These results support the ``Unified model of AGN'' that type 1 and
type 2 Seyfert galaxies are essentially the same, and the differences 
can be explained by the presence or absence of optically and
geometrically thick molecular torus in our line of sight.

Mrk 1210 is one of the nearby Seyfert 2 galaxies at a redshift of 0.0135,
and is well-known as a strong H$_2$O maser emitter. An X-ray study of Mrk
1210 was first performed with ASCA in 1995, and it has been found
to exhibit a complex spectral feature.
The spectrum consists of two components: a soft
component and a flat hard component, with a neutral Fe-K line
(\cite{Awaki}). 
However, poor photon statistics and a lack of information
above 10 keV prevented them from constraining the spectral shape
in detail. 
Awaki et al (2000) suggested two possibilities for the
origin of the hard component. 
One possibility is that the hard component is
a heavily absorbed direct nuclear emission. 
In that case, the absorption column density and absorption-corrected 
X-ray luminosity are $N_{\rm H}\sim2.0\times 10^{23}  $cm$^{-2}$ 
and $L_{\rm ac}\sim 2.3 \times 10^{42}$ erg s$^{-1}$, respectively. 
The other is that the nuclear emission is completely obscured in
the ASCA band, and they detected only the scattered light, where 
the intrinsic X-ray luminosity is estimated to be
$L_{\rm int}\sim10^{43}$ erg s$^{-1}$. 
Awaki et al. (2000) also claimed that the latter case is preferred, 
based on the correlation between FIR and X-ray for other
Compton-thick Seyfert 2 galaxies.

Mrk 1210 is a valuable object for X-ray studies, since we can obtain
important information of on the nuclear region, combining the X-ray data with the
maser data. However, as described above, previous X-ray observations
were not sufficient due to the limited energy band. 
In order to constrain the spectral shape
unambiguously, we performed an X-ray observation with BeppoSAX by utilizing its
wide energy band up to 100 keV (\cite{Boella}).

By accident, XMM-Newton observed Mrk 1210 on just the same day as our
BeppoSAX observation.
The result has been reported by Guainazzi et al. (2002); 
the spectral shape is dramatically
different between the ASCA and XMM-Newton observations.
The flux of the hard component in the XMM-Newton observation is about 
5-times as luminous as that of the ASCA, and an absorption cut-off
is clearly detected around 4 keV.
Following these properties, Guainazzi et al. (2002) called this object 
 ``the Phoenix galaxy'', and suggested that this nucleus made a 
transition from the reflection-dominated quiescence 
to the X-ray luminous state.
Such phenomena were also reported for NGC 6300 (\cite{Guainazzi3}) and NGC 2992
(\cite{Gilli}).
However, the alternative scenario of a large change of the absorption 
column density by a factor of $>5$ cannot be ruled out.
A variability of the absorption column density of Seyfert 2 galaxies
has been reported by Risaliti, Elvis, and Nicastro. (2002), 
mainly for Compton-thin objects.
One possible example for Compton-thick objects is M 51; the BeppoSAX
detected a strongly absorbed nuclear emission above 10 keV (\cite{Fukazawa}), while an unabsorbed nuclear emission was observed with
Ginga (\cite{Makishima}).
In this case, a dramatic change in the absorption from $>10^{24}$ cm$^{-2}$
to $<10^{21}$ cm$^{-2}$ is required.

In the case of Mrk 1210, both the ASCA and XMM-Newton observations were
limited to the $<10$ keV band.
The key spectral feature to extract possible evidence of variability
of nuclear emission is a reflection component.
Although the response of the reflector is smeared with a time
scale of several years, any sign of
very high amplitude of the nuclear variability would be seen in the
reflection component.
In order to investigate the possibility of an absorption change for Mrk
1210, we here reported the results of the BeppoSAX observation.

\section{Observation and Data Reduction}

We observed Mrk 1210 with BeppoSAX on 2001 May 5 (PI; N. Iyomoto).
The exposure time is 31.6 ks for the Low-Energy Concentrator/Spectrometer (LECS,
\cite{Parmar}), 
95 ks for the Medium Energy Concentrator/Spectormeter 2 \& 3 (MECS2+3, \cite{Boella}), and 41 ks for the
Phoswich Detection System (PDS, \cite{Frontera}). 
One bright point source associated with the nuclear 
region was detected in the LECS and MECS X-ray image.   
We accumulated the LECS and MECS data within 5'
radius centered on the nucleus to construct the spectra. 
On the other hand,
we made use of the background-subtracted spectrum 
prepared by BeppoSAX Science Data Center (SDC) for the PDS.
We applied the response file in
1998 November for the MECS and in 1999 December for the LECS, both of
which are supplied by SDC.  
For background subtraction, we also used the standard blank sky data of
1997 May for the MECS and 1999 November for the LECS, 
and we integrated the background
spectrum from the same region as the on-source spectrum for each
detector. 
We rebinned the spectra to contain more than 20 counts in each bin to apply
the $\chi^2$ method for a spectral fitting.

\section{Results}

\subsection{Spectral Analysis}

\begin{figure}[htbp]
\begin{minipage}{8.cm}
\resizebox{7.5cm}{!}{\includegraphics{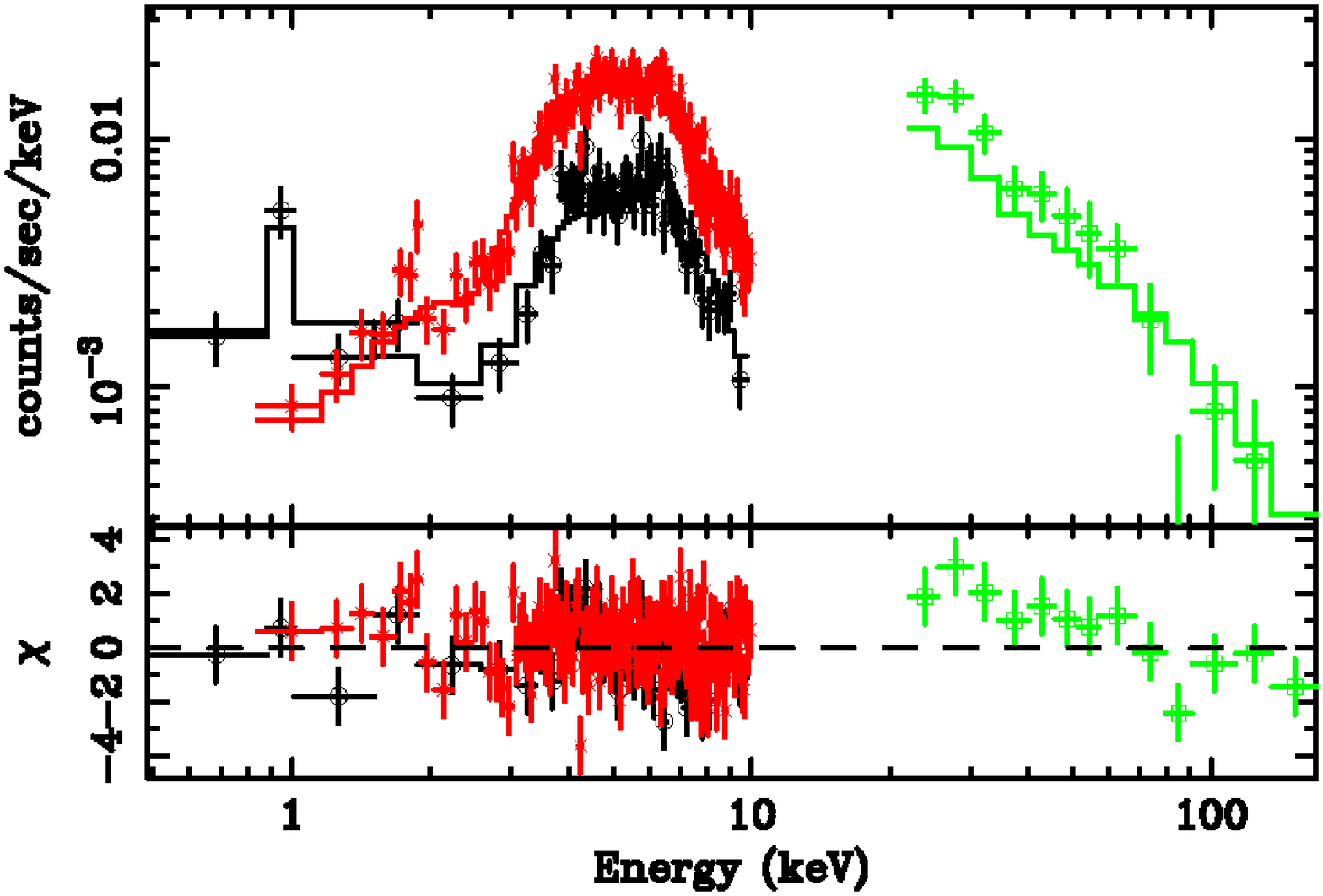}}
\end{minipage}
\begin{minipage}{8.0cm}
\resizebox{7.5cm}{!}{\includegraphics{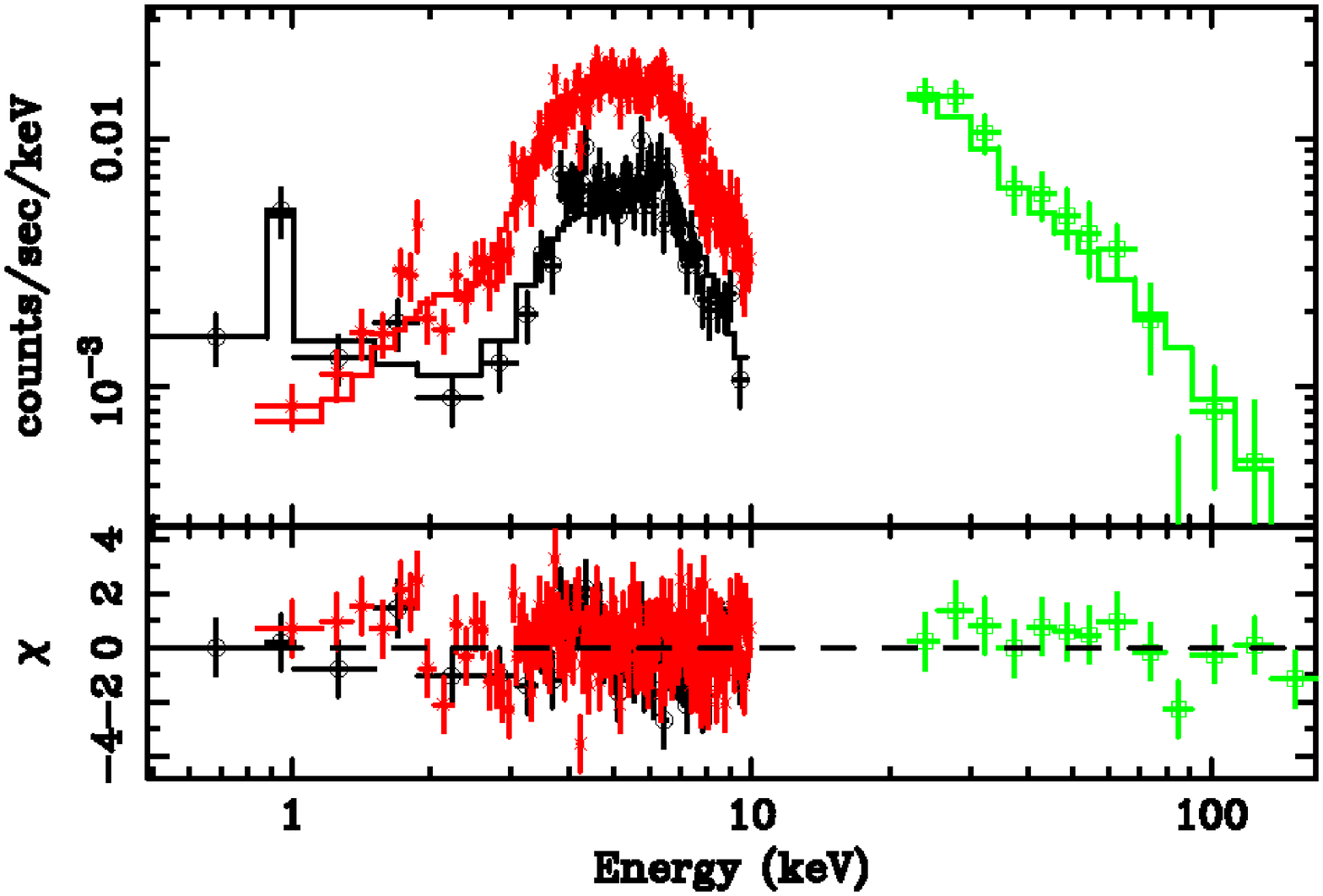}}
\end{minipage}
\vspace*{0.5cm}
\caption{X-ray spectra of Mrk 1210 obtained with the LECS (circle),
 MECS23 (asterisk), and PDS (square). The solid line represents the
 best-fit model 1 for the left and the model 2 for the right, 
as shown in table 1.}
\label{saxspec}
\end{figure}

The spectra of the LECS, MECS2+3, and PDS detectors are shown in
figure \ref{saxspec}.
The spectral shape is quite different from that of ASCA,
and two components are clearly seen; a heavily
absorbed bright continuum in the hard band and a faint soft component. 
Such a heavily absorbed hard component was not clearly detected with
ASCA, 
and it is so luminous that it can be detected with the PDS up to 100 keV. 
We also detect a Fe-K line structure around 6.4 keV. 
The flux and line structures around 1 keV are 
in good agreement with the ASCA and XMM-Newton results.

We then performed a spectral fitting by employing an absorbed power-law
 model to express the heavily absorbed hard continuum.
We also added another power-law component with the same photon index as
 mentioned above
to express the soft component, following Guainazzi et al. (2002), 
who claimed that the
soft component is a reflection by warm material, based on the 
center energy of the Si-K line.
Since the BeppoSAX data could not resolve the Si-K line, we included only
the line at 0.93 keV.
The parameters of the soft component are fixed to those obtained by 
XMM-Newton, since the soft component is too faint to determine them.
We also included a Gaussian model to represent the Fe-K line, 
and a Galactic absorption of $N_{\rm H}=3.53\times10^{20}$ cm$^{-2}$ 
(\cite{Dickey}).
Table \ref{fitpar} shows the obtained best-fit parameters.
The photon index of the power-law became 1.61$^{+0.05}_{-0.07}$, and the hard component suffered
heavy absorption of $N_{\rm H} = 1.83^{+0.08}_{-0.12}\times
10^{23}$ cm$^{-2}$. 
The Fe-K line is significant with an equivalent
width of 108$^{+50}_{-65}$ eV, and 
energy of the Fe-K line center is consistent with that of the neutral
iron in the rest frame of Mrk 1210.
The PDS spectrum is well-fitted by
the power-law model with the same parameters as the LECS or MECS2+3. 
The observed 2--10 keV X-ray flux
and absorption-corrected X-ray luminosity became  $F_{\rm X} = 9.3 \times
10^{-12}$ erg s$^{-1}$ cm$^{-2}$ and $L_{\rm X} = 1.6 \times 10^{43}$
erg s$^{-1}$, respectively. 
We here assumed the Hubble constant being $H_0 = 50$ km s$^{-1}$ Mpc$^{-1}$ 
to calculate the X-ray luminosity. 
The observed flux is higher by a factor $>6$ than that of the ASCA data.
These results are in good agreement with those of the XMM-Newton observation
 (\cite{Guainazzi}).

Next, we tried to constrain the reflection component, which has often
been detected for Seyfert 2 galaxies (e.g. \cite{Guainazzi3}).
In figure 1, we see some hint of a convex spectral feature around 30--50
keV in the residual.
Furthermore, the photon index of 1.61 is relatively small compared 
with other Seyfert 2 galaxies, indicating an additional hard component.
Then, we estimated a Compton reflection component by including the
PEXRAV model (\cite{Magdziarz}) in the XSPEC ver 11.2.0. 
We assumed the solar abundance for the reflector, and
30 degrees for the inclination angle.
Since the high-energy cut-off cannot be constrained well, we fixed it to 
10 MeV.
Thus, the results are summarized in table \ref{fitpar}.
The reflection component is significantly required to explain the
convex structure at around 30--50 keV, as shown in figure \ref{saxspec}.
The power-law photon index becomes 1.86$^{+0.09}_{-0.10}$, 
typical values of Seyfert galaxies.
The equivalent width of the
Fe-K line against the reflection component is $523^{+547}_{-443}$ eV, 
similar to that of the ASCA observation.

\begin{figure}[htbp]
\begin{center}
\resizebox{14cm}{!}{\includegraphics{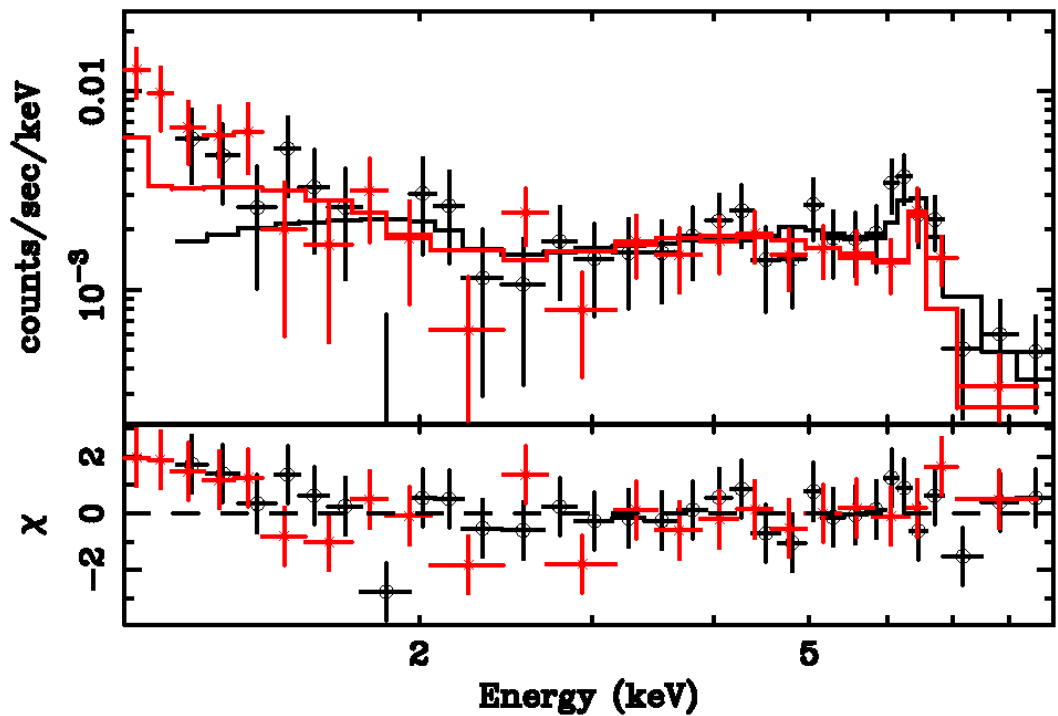}}
\caption{Comparison of the ASCA spectra with the best-fit 
reflection component for the BeppoSAX spectra.}
\label{ascaspec}
\end{center}
\end{figure}

We are interested in whether the ASCA hard component is 
represented by only a reflection component.
We compared the above best-fit model with the ASCA data, 
by excluding the absorbed bright power-law component.
As shown in figure \ref{ascaspec},
the model is in good agreement with the ASCA data with a reduced
$\chi^2$ of 52/43.
Thus, the flat hard component observed with ASCA is likely to be a reflection
component, indicating that the reflection flux did
not vary between the ASCA and BeppoSAX observations in a time interval of
6 years.
In addition, we tried to include a bright power-law
component of the same normalization with BeppoSAX, but 
strongly absorbed with $>1.5\times10^{24}$ cm$^{-2}$.
This model also represents the ASCA spectra with the same $\chi^2$
value as before, including the strongly absorbed component, 
claiming that we cannot
completely distinguish whether the direct nuclear emission was blocked
or intrinsically very faint below 10 keV in the ASCA observation.

\subsection{Timing Analysis}

\begin{figure}[htbp]
\begin{minipage}{8.0cm}
\begin{center}
\resizebox{7.5cm}{!}{\includegraphics{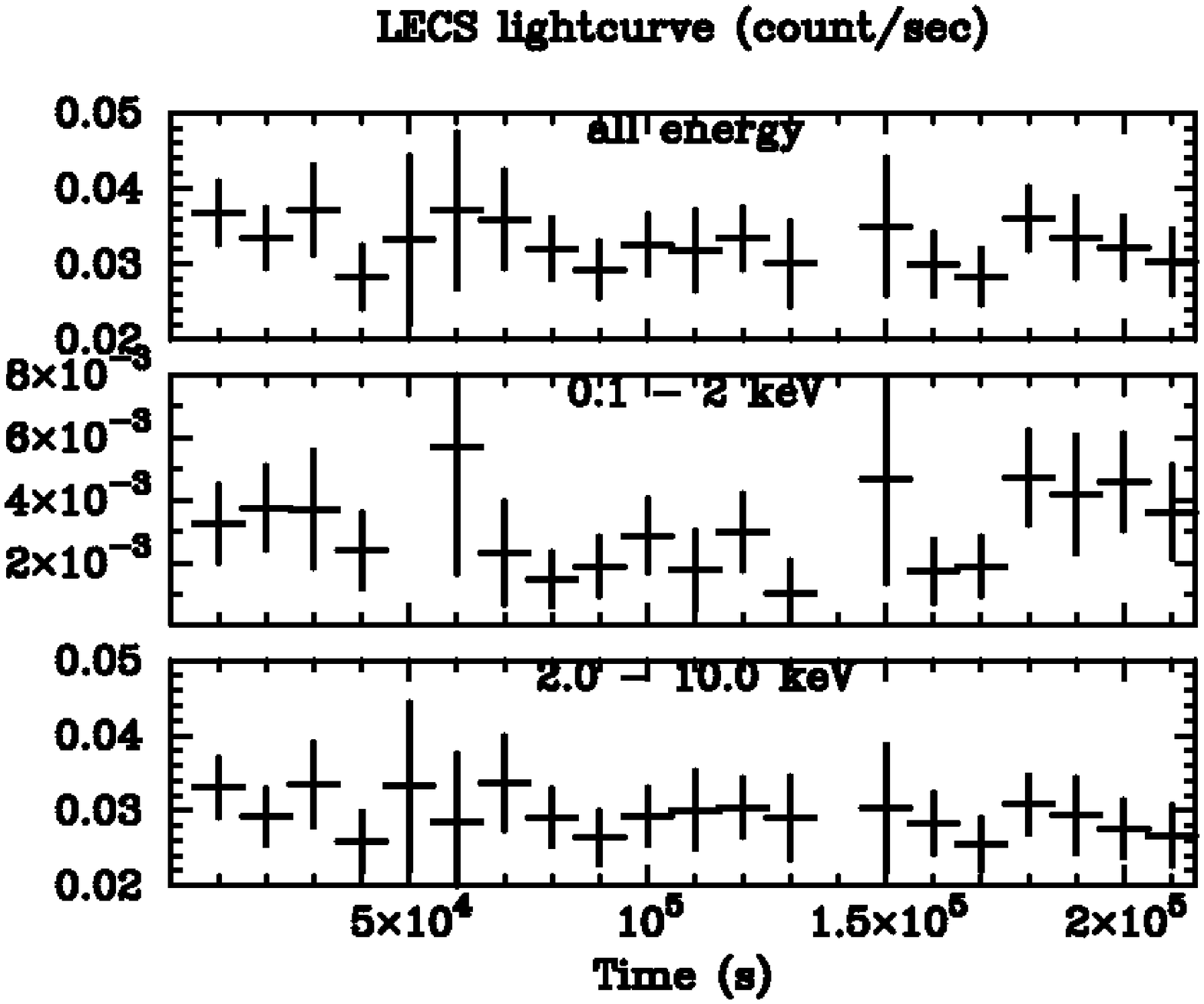}}
\end{center}
\end{minipage}
\begin{minipage}{8.0cm}
\begin{center}
\resizebox{7.5cm}{!}{\includegraphics{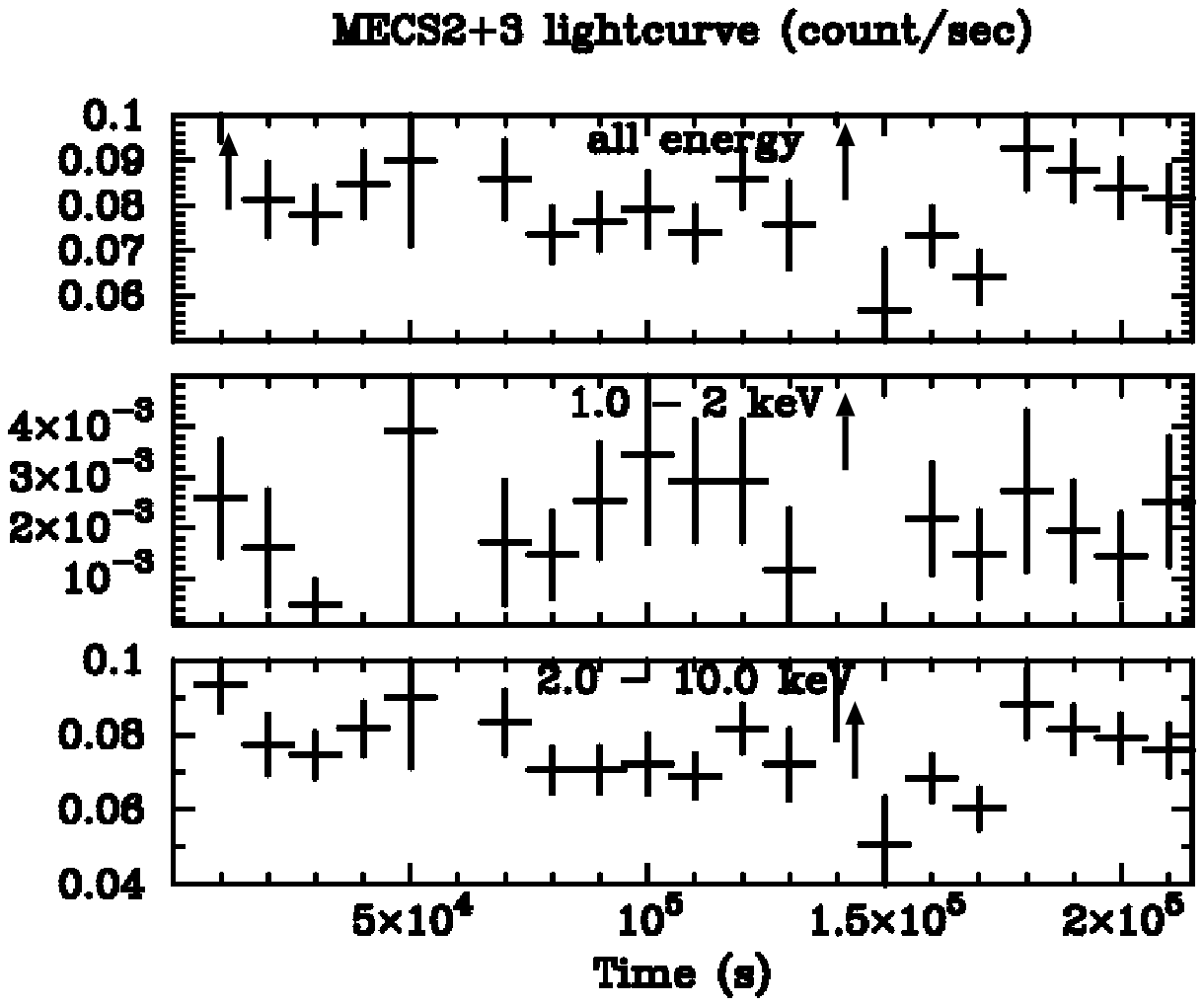}}
\end{center}
\end{minipage}
\caption{X-ray lightcurves of the LECS (left) and the MECS (right) in
 the three energy bands.}
\label{lightcur}
\end{figure}

Figure \ref{lightcur} shows X-ray lightcurves of the LECS and MECS2+3,
where the time bin width is 10000 s. Unfortunately, there is data
loss around 60000 s and 150000 s in both LECS and MECS2+3 lightcurves.
When we did not include the periods at 52400 s - 68500 s and 138000 s -
154000 s, the reduced $\chi^2$ for
assuming the constant flux becomes 1.57 in the 2--10 keV band. 
Such a reduced $\chi^2$ implies variability in the MECS 
lightcurves around the time of 160000 s.
On the other hand, we found no variability in the MECS soft band and the
LECS lightcurves.
Therefore, the time variability is possibly caused by only the hard
component, although the variability is not so significant.
Since the hard component is a direct nuclear emission, as described
before, 
the possible variability in the MECS data implies a change of nuclear
activity on a time scale of 10000 s. 
Such a sub-day time variability of the Seyfert 2 nucleus is also found in
a similar object, NGC 6300 (\cite{Guainazzi3}).
Since the XMM-Newton observation was performed at only the first $\sim$10000
sec of the BeppoSAX observation, we cannot confirm the consistency.

\begin{table}[htbp]
\begin{center}
\caption{The best-fit parameters of spectral fittings.}
\label{fitpar}
\begin{tabular}{lllllll}
\hline
\hline
Model&P-L&Absorption&\multicolumn{2}{c}{Fe-K line}&\small {Refl
 Comp}&\\
&$\Gamma~^\ast$&$N_{\rm H}~^\dagger$&$E_c~^\ddagger$&E.W.$~^\S$ &$R~^\parallel$&$\chi^2$/d.o.f.\\
\hline
1 & 1.61$^{+0.05}_{-0.07}$ &1.83$^{+0.08}_{-0.12}$& 6.39$^{+0.09}_{-0.09}$ & $108^{+50}_{-65}$ & &291/210\\
2 & 1.86$^{+0.09}_{-0.10}$ &2.04$^{+0.16}_{-0.14}$& 6.39$^{+0.10}_{-0.10}$ & $523^{+547}_{-443}$ &
 1.86$^{+0.93}_{-0.76}$ &  265/208\\

\hline
\multicolumn{7}{l}{\scriptsize{Errors are at 3$\sigma$ level.}}\\
\multicolumn{7}{l}{\scriptsize{model 1; power-law + absorbed power-law, model 2; power-law + absorbed power-law + reflection(PEXRAV).}}\\
\multicolumn{7}{l}{\scriptsize{$\ast$: photon index of the power-law model.}}\\
\multicolumn{7}{l}{\scriptsize{$\dagger$: absorption column density ($\mathrm{10^{23} cm^{-2}}$).}}\\
\multicolumn{7}{l}{\scriptsize{$\ddagger$: line center energy (keV).}}\\
\multicolumn{7}{l}{\scriptsize{$\S$: line equivalent width (eV) against
 the powerlaw component or the reflection component for the model 1 or 2
 , respectively.}}\\
\multicolumn{7}{l}{\scriptsize{$||$: scaling factor of reflection (R = 1 for isotropic source above disc).}}\\
\end{tabular}
\end{center}
\end{table}

\section{Discussion}

\subsection{Long-Term Variability of Spectral Shape}

In our BeppoSAX observation of Mrk 1210, we clearly detected the absorbed
direct emission from the nucleus in the hard X-ray band up to 100 keV.
The spectral shape and flux are found to be quite
different from those of the ASCA data, and consistent with those of the
XMM-Newton data below 10 keV.
Thanks to the PDS data, our observation was able to find that the reflection
component is required significantly.
This reflection component reproduces the ASCA spectra very well. 
Based on these results, we discuss the time variability of Mrk
1210 between the ASCA and BeppoSAX/Newton observation.

We here consider three possibilities.
The first is that the hard component in the ASCA
data is  heavily absorbed nuclear emission, as suggested by Awaki et
al. (2000).
In this case, the absorption column density is
found to be almost the same as $N_{\rm H} \sim 2.0 \times 10^{23}$ 
cm$^{-2}$ between the two observations. 
Then, the absorption-corrected X-ray luminosity in the 2 --10 keV
becomes $L_{\rm ac} = 1.62^{+0.13}_{-0.12}\times10^{43}$ 
erg s$^{-1}$ in the BeppoSAX observation, which
is higher by a factor of $>$6 than that in the ASCA
observation. 
Therefore, it is required that the large spectral difference between ASCA and
BeppoSAX is due to the luminosity change of the nucleus.
There are few Seyfert 2 galaxies that exhibit such a large change of the
X-ray luminosity; for example, Mrk 3 is a famous one (\cite{Iwasawa}).
However, we think that a luminosity change is not likely, 
because the ASCA spectra
can be modeled well with the reflection component derived from the
BeppoSAX data, and thus indicates a constant flux from the nucleus.

The second possibility is that the hard component of the ASCA data is an
echo, in other words, 
a reflection of the nuclear emission, which itself was very faint in the
ASCA observation,
as claimed by Guainazzi et al. (2002), in analogy with NGC 4051
(\cite{Guainazzi2}) and NGC 6300 (\cite{Guainazzi3}), whose nuclear
emission certainly disappeared even above 10 keV.
However, we cannot accept it straightforwardly 
since the ASCA data did not confirm the flux above 10 keV.
This case is similar to the first possibility in assuming that the
flux of nuclear emission varied dramatically.
The flux of the reflection
component should also have followed that of the nuclear direct emission 
with a time delay of several years.
If the nuclear emission in the ASCA observation was very different
from that in the BeppoSAX observation, 
the reflection component was thought to also be different. 
However, our result shows that the flux of reflection component is almost the
same between ASCA and BeppoSAX.
Therefore, we consider this not to be preferable. 

Accordingly, we suggest the third possibility that the hard 
component of the ASCA data is a reflection of the nuclear emission, 
which itself is intrinsically bright, but completely obscured below 10 keV. 
This is indicated by
the correlation between the X-ray and far-infrared luminosity for Seyfert 2
galaxies (\cite{Awaki}), and 
the intrinsic X-ray luminosity 
is predicted to be $L_{\rm int} \sim 10^{43}$ erg s$^{-1}$, which is almost the
same as the absorption-corrected luminosity obtained by BeppoSAX.
This possibility gives rise to that a large absorption column density 
of $N_{\rm H} > 10^{24}$ cm$^{-2}$ blocked the direct emission in the ASCA
observation, and that the absorption decreased down to $N_{\rm H}\sim10^{23}$ 
cm$^{-2}$ in the BeppoSAX observation.
Risaliti, Elvis, and Nicastro. (2002) reported that many Seyfert 2 galaxies have shown a
change of absorption column density, but
few objects have exhibited such a large change by an order of magnitude 
as Mrk 1210. 
M 51 is one of such rare cases (\cite{Fukazawa}).

In summary, the differences in the spectra of Mrk 1210 between ASCA and
BeppoSAX is likely to be due to the dramatic variability of 
either the X-ray luminosity of the nuclear emission or the absorption column density 
of the molecular torus on a time scale of 6 years. 
As above, we claim that the latter is preferable,
although we cannot rule out the former because of the limitation of
the ASCA energy band. 
Therefore, we need observations with an instrument that has a
wider energy band up to a few hundreds keV, such as Integral or Astro-E2 HXD.
In any case, it can be said that Mrk 1210 is a very important object to 
investigate the nuclear structure of Seyfert 2 galaxies,
since it is a rare Seyfert 2 galaxy that shows a large
variability.

\subsection{Molecular Torus}

Considering the BeppoSAX results, we can impose some constrains on the 
geometry of the molecular torus around the nucleus of Mrk 1210.
First of all, we estimate how distant the molecular torus exists from the
nucleus, based on the maser results. 
Mrk 1210 is known to associate a bright H$_2$O maser emission. 
Let us assume roughly that the maser emission originates at the most
inner radius of the molecular torus. 
The maser was observed at 4214 km s$^{-1}$ against a systemic 
velocity of 4046 km s$^{-1}$ of the Mrk 1210 galaxy (\cite{Braatz}), and thus the maser-emission region rotates at
a velocity of 168 km s$^{-1}$. 
The mass of the central blackhole is assumed to be $10^7 M_{\odot}$, based
on the optical bulge luminosity of $\sim10^{10} L_{\odot}$, 
a mass-to-luminosity ratio of $M/L\sim6$, 
and a typical blackhole-to-bulge mass ratio of $\sim$0.002 
(Kormendy, Richstone 1995).
Considering the Keplarian motion around the nuclear blackhole, 
it is estimated that the molecular torus 
exists at less than 1 pc from the nucleus. 
This is consistent with the value estimated for other 
Seyfert 2 galaxies (\cite{Risaliti2}), and thus the overall dimension
of torus is on the order of 1 pc.

Taking this constraint into account, we examine the two possibilities
described in subsection 4.1. 
In the case of the variability of the X-ray luminosity, 
the absorption column density of the torus has not changed in 6 years. 
Therefore, we can say that the size of the molecular torus is more than 2
pc, or the molecular torus is homogeneous. 
While in the other case, we can constrain the size of
the molecular torus to be less than 0.001 pc, considering the rotation
velocity of the torus and the variability time scale. 
Since the latter case is, we think, preferable, 
it is suggested that the torus is not homogeneous, 
but like a blob structure.

\vspace*{0.5cm}
The authors thank Dr. H. Awaki for a careful reading of the manuscript and 
many helpful comments.
The authors are also grateful to the BeppoSAX team for
their help in the spacecraft operation, calibration, and data analysis.

\end{document}